\documentclass[aps,twocolumn,showpacs, showkeys,nofootinbib,floatfix]{revtex4-1}

\usepackage{amsmath}
\usepackage{amsfonts}
\usepackage{txfonts}
\usepackage{graphicx}
\usepackage{dcolumn}
\usepackage{bbm}
\usepackage{amssymb}
\usepackage{latexsym}
\usepackage{titlesec}
\usepackage[colorlinks=true, linkcolor=red, citecolor=blue]{hyperref}
\usepackage{longtable}

\begin{document}

\title{Gravitational Waves: A Test for Modified Gravity}

\author{Lixin Xu}
\email{Corresponding author. \\
lxxu@dlut.edu.cn}

\affiliation{Institute of Theoretical Physics, School of Physics \& Optoelectronic Technology,
Dalian University of Technology, Dalian 116024, People's Republic of China}

\affiliation{State Key Laboratory of Theoretical Physics, Institute of Theoretical Physics, Chinese Academy of Sciences, Beijing 100190, People's Republic of China}

\begin{abstract}
In a modified gravity theory, the propagation equation of gravitational waves will be presented in a non-standard way. Therefore this tenor mode perturbation of time-space, as a complement to the scalar mode perturbation, provides a unique character distinguishing modified gravity from general relativity. To avoid the model-dependent issue, in this paper, we propose a parametrised modification to the propagation of gravitational waves. We show the effects on the angular power spectrum of cosmic microwave background radiation due to the parametrised modification and its degeneracy to the tensor mode power spectrum index $n_t$ and its running $\alpha_t$. At last, we report the current status on the detection of modified gravity through the currently available cosmic observations. Our results show no significant deviation to general relativity.    
\end{abstract}


\maketitle

\section{Introduction}

Commonly realising a late time accelerated expansion of our Universe needs a modification to general relativity (GR) at large scales dubbed as modified gravity (MG), or an addition of exotic energy component named dark energy (DE). However, these realisations are totally different in nature. Finding out a modification to GR at large scales means the discovery of new gravity theory beyond GR. Confirming the existence of DE implies the discovery of new particle(s) beyond the standard particle physics model. The issue is how to distinguish MG theorise from DE models. It is believed that cosmic observations provide the finally experimental judgement in addition to the fundamental physics theory.   

Due to the existence of a great diversity of MG theories and DE models, it is almost impossible to test every model. One possibility is finding out a general formalism which can grasp the main characters of MG theories and DE models, for example a feasibly parametrised MG and DE model that are consistent to cosmic observations, but may be independent on any concrete fundamental physics. Based on this spirit, a parametrised modification to GR in the scalar mode perturbation was studied in the literature, see \cite{ref:BZ,ref:MGCAMB,ref:mugammaforms} and reference therein for examples. The modification was mainly focused on the Poisson equation and the slip of the Newtonian potentials of $\Phi$ and $\Psi$, but keeping the background evolution to a standard $\Lambda$CDM cosmology, say 
\begin{eqnarray} 
k^2\Psi&=&- \mu(k,a) 4\pi G a^2\left[ \rho \Delta+3(\rho+P)\sigma\right],\label{eq:MGpoisson}\\
k^2[\Phi-\gamma(k,a)\Psi]&=&\mu(k,a) 12\pi G a^2(\rho+P) \sigma,\label{eq:MGslip}
\end{eqnarray} 
in Fourier $k$-space as an example borrowed from Ref. \cite{ref:MGCAMB}, where $\Delta=\delta+3\mathcal{H}(1+w)\theta/k^2$ is the gauge-invariant overdensity and $\delta\equiv \delta\rho/\rho$ is the overdensity of energy component $\rho$; $\mathcal{H}\equiv \dot{a}/a$ is the conformal Hubble parameter, here the dot $\dot{}$ denotes the derivative with respect to the conformal time $\tau$, and $a$ is the scale factor; $w$ is the equation of state of energy component $\rho$; $\theta$ is divergence of the velocity perturbation, i.e. the peculiar velocity and $\sigma$ is the anisotropic stress. $\Psi$ and $\Phi$ are the Newtonian potentials in the conformal Newtonian gauge
\begin{equation}
ds^2=a(\tau)^2\left\{-(1+2\Psi)d\tau^2+\left[(1-2\Phi)\delta_{ij}+2h^{T}_{ij}\right]dx^{i}dx^{j}\right\},
\end{equation}
where $h^{T}_{ij}$ is a traceless ($h^{Ti}_{i}=0$), divergence-free ($\nabla^{i}h^{T}_{ij}$), symmetric ($h^{T}_{ij}=h^{T}_{ji}$) tensor field. The two $\mu(k,a)$ and $\gamma(k,a)$ are scale and time dependent functions encoding any modification to gravity theory in scalar mode. Note that GR is recovered in the $\mu=\gamma=1$ limit. When considering the locality, general covariance and the quasi-static approximation, the physically acceptable forms of $\mu(k,a)$ and $\gamma(k,a)$ should be the ratios of polynomials in even $k$, with numerator of $\mu$ set by the denominator of $\gamma$ \cite{ref:mugammaforms}. Therefore, for the scalar part modification, one obtains \cite{ref:mugammaforms}, see also in Refs \cite{ref:Felice2011,ref:Solomon2014,ref:Konnig2014}
\begin{eqnarray}
\gamma(k,a)=\frac{p_{1}(a)+p_{2}(a)k^2}{1+p_{3}(a)k^2},\quad \mu(k,a)=\frac{1+p_{3}(a)k^2}{p_{4}(a)+p_{5}(a)k^2},
\end{eqnarray} 
where $p_{i}(a),i=1...5$ are functions of $a$. The $\mu(k,a)$ and $\gamma(k,a)$ can be fixed for a specific MG model, see Refs. \cite{ref:Bellini2014,ref:Baker2014} for examples. However, the {\it Planck} 2015 DE and MG paper has shown that the scale dependence of $\mu$ and $\gamma$ does not lead to a significantly small $\chi^2$ with respect to the scale-independent case \cite{ref:Planck2015DEMG}. This may come from the insufficiency of the large scale structure information. Therefore in this paper, we only consider the scale-independent forms. After eliminating the scale dependence, the final form for $\mu$ and $\gamma$ should be
 \begin{eqnarray}
\gamma(k,a)=p_{\gamma}(a),\quad \mu(k,a)=\frac{1}{p_{\mu}(a)},\label{eq:MGscalar}
\end{eqnarray}    
for explicity we propose $p_{i}(a)=\lambda_{i} a^{s_{i}}, (i=\gamma,\mu)$ as a woking example. In phenomena, it can be a function of $w_{de}(a)$ and $\Omega_{de}(a)$ effectively, say $p_i(a)=\lambda_{s_{i}} w_{de}(a) \Omega_{de}(a)^{s_{i}}$ etc. 

The tensor mode perturbation, as a complement to the scalar mode perturbation, should also be altered in a modified gravity theory. In GR, the propagation of gravitational waves in Fourier $k$-space is written as
\begin{equation}
\ddot{h}_{ij}^{T}+2\mathcal{H}\dot{h}_{ij}^{T}+c^2_T(k^2+2K)h_{ij}^{T}=8\pi G a^2 \Pi_{ij},
\end{equation}
where the transverse-traceless tensor $\Pi_{ij}$ is the anisotropic part of the stress tensor, $K$ the three-dimensional curvature ($K=0$ is adopted in this paper) and $c^2_T$ is the square of the speed of gravitational waves.  In the literature \cite{ref:MGtensor1,ref:MGtensor2,ref:MGtensor3,ref:MGtensor4,ref:MGtensor5}, a parameterised modification to the tensor mode perturbation was proposed  recently. In general, the propagation of gravitational waves is modified due to the interaction between new degree of freedom (introduced for providing late time accelerated expansion of our Universe) and curvature or metric \cite{ref:MGtensor1},
\begin{eqnarray}   
\ddot{h}_{ij}^{T}+2\mathcal{H}\left[1+\frac{\chi(k,a)}{2}\right]\dot{h}_{ij}^{T}&+&c^2_T(k^2+2K)h_{ij}^{T}+a^2m_{g}^2h_{ij}^{T}\nonumber\\
&=&8\pi G a^2\Gamma(k,a) S_{ij},\label{eq:MGtensor}
\end{eqnarray}  
where $\chi\equiv\mathcal{H}^{-1}(d\ln M^2_{\ast}/dt)$ describes the running rate of the effective {\it Planck mass} $M_{\ast}$, $m_{g}$ is the mass of graviton in a massive gravity theory. And the transverse-traceless tensor $S_{ij}$ is the source term for the gravitational waves. The form of this source term $S_{ij}$ depends on MG theories or the properties of matter fluid \cite{ref:MGtensor1,ref:MGDEKunz}. It is very important to note that when the source term $S_{ij}$ comes from the anisotropic stress of matter fluid, the homogenous part of Eq. (\ref{eq:MGtensor}) will never be modified \cite{ref:MGtensor1}. Therefore the anisotropic stress plays the role of a signature of nonstandard propagation of gravitational waves. It also implies that any significant $\xi(k,a)\neq 0$ or $m_g\neq 0$ signals the detection of MG which cannot be disguised by DE. The $\Gamma(k,a)$ term modification would be related to the $\xi(k,a)$ term for a specific MG model,  for example in $f(R)$ gravity the propagation of gravitational waves is given by \cite{ref:Nohfr}
\begin{equation}
\ddot{h}_{ij}^{T}+2\mathcal{H}\left(1+\frac{1}{2}\frac{d\ln F}{d\ln a}\right)\dot{h}_{ij}^{T}+c^2_T(k^2+2K)h_{ij}^{T}=\frac{8\pi G a^2 \Pi_{ij}}{F},\label{eq:FRtensor}
\end{equation}
where $F\equiv d f/dR$. Sometimes $c^2_{T}$ will also deviate from the speed of light in a MG model, for examples the scalar-tensor and Einstein-aether models as shown in Ref. \cite{ref:MGtensor1}. In Ref. \cite{ref:MGmeature}, the speed of the cosmological gravitational waves was constrained by using the {\it Planck} 2013 and BICEP2 data sets, where no significant deviation from the standard values $c^2_{T}=1$ was probed: $c^2_{T}=1.30\pm 0.79$. Therefore in this paper, we will fix $c^2_{T}$ to its standard value. But we also will show the possible degeneracy between the $c^2_{T}$ and the $\chi(k,a)$ term in Section \ref{sec:cmbttbb}. Motivated by the modification coming from $f(R)$ gravity, say Eq. (\ref{eq:FRtensor}), one can proposes a modified equation in the following form
\begin{equation}
\ddot{h}_{ij}^{T}+2\xi(k,a)\mathcal{H}\dot{h}_{ij}^{T}+c^2_{T}(k^2+2K)h_{ij}^{T}=8\pi G \mu(k,a)\Pi_{ij},\label{eq:tensorp}
\end{equation}    
where $\xi(k,a)$ and $\mu(k,a)$ are two functions encoding a modified gravity theory. Since the $\xi(k,a)$ characterise the running rate of the effective {\it Placnk mass} which should be scale-independent, therefore we assume
\begin{eqnarray}
\xi(k,a)=p_t(a).\label{eq:MGxia}
\end{eqnarray}  
And we will take $p_t(a)=\lambda_t a^{s_{t}}$ as a working example. In phenomena, it can be a function of $w_{de}(a)$ and $\Omega_{de}(a)$ effectively, say $p_t(a)=\lambda_t w_{de}(a) \Omega_{de}(a)^{t}$ etc. 

Recently, the Background Imaging of Cosmic Extragalactic Polarization (BICEP2) experiment \cite{ref:BICEP21,ref:BICEP22} has detected the B-modes of polarization in the cosmic microwave background, where the tensor-to-scalar ratio $r=0.20^{+0.07}_{-0.05}$ with $r=0$ disfavored at $7.0\sigma$ of the lensed-$\Lambda$CDM model was found. Recently, {\it Planck 2015} released the polarisation result and didn't find significant signal of the primordial gravitational waves. However the {\it Planck 2015} TT, EE, and TE data are not released now. So in this paper, we still use the {\it Plank} 2013 data points. But our analysis on the CMB TT and BB power spectrum doesn't depend on the data points. 

This paper is structured as follows. At first, in Section \ref{sec:cmbttbb}, we show the effects on CMB TT and BB power spectrum due to the parametrised modification to GR along with Eq. (\ref{eq:MGscalar}), Eq. (\ref{eq:tensorp}) and Eq. (\ref{eq:MGxia}). To confirm these effects purely coming from the parametrised modification, we also test the possible degeneracy to the tensor spectrum index $n_t$ and its running $\alpha_t=d n_t/d\ln k$. We report the current probe of MG by performing a global Markov chain Monte Carlo (MCMC) analysis in Section \ref{sec:results}. Section \ref{sec:conclusion} is the conclusion.  
 
\section{Effects on the CMB TT and BB Power Spectrum} \label{sec:cmbttbb}

To study the effects on the CMB TT and BB power spectrum arising from the parameterised modification to GR, we modified the {\bf MGCAMB} code \cite{ref:MGCAMB} to include the tensor perturbation equation as shown in Eq. (\ref{eq:tensorp})-(\ref{eq:MGxia}). In the {\bf MGCAMB} code, the CMB TT source term from the scalar mode perturbation in terms of the synchronous gauge variables is given by \cite{Zaldarriaga:1996xe}
\begin{eqnarray}
S^{(S)}_T (k,\tau)&=&g\left(\Delta_{T0} +2 \dot{\alpha}+{\dot{v_b} \over k}+{\Pi \over 4 }+{3\ddot{\Pi}\over 4k^2 }\right)\nonumber\\
&+&e^{-\kappa}(\dot{\eta}+\ddot{\alpha})+\dot{g}\left(\alpha+{v_b \over k}+{3\dot{\Pi}\over {2}k^2 }\right)+{3 \ddot{g}\Pi \over 4k^2},\label{eq:Tsource}
\end{eqnarray}
where $\kappa$ is the optical depth, $g$ is the visibility function, $\Pi = \Delta^T_2+\Delta^P_2+\Delta^P_0$ and $\Delta^T_\ell (\Delta^P_\ell)$ are the $\ell$'th moments of $\Delta^T(\Delta^P)$ in term of Legendre polynomials \cite{Zaldarriaga:1996xe}; the $\alpha$ term is changed to \cite{ref:MGCAMB} 
\begin{equation}
\alpha =  \left\lbrace \eta + \frac{\mu 8\pi G a^2}{2k^2} \left[ \gamma \rho \Delta + 3(\gamma -1) (\rho + P) \sigma \right]  \right\rbrace / \mathcal{H} ,\label{eq:alpha}
\end{equation} 
in terms of $\mu$ and $\gamma$ for the parametrised MG; the integrated Sachs-Wolfe (ISW) effect term $e^{-\kappa}(\dot{\eta}+\ddot{\alpha})$, in (\ref{eq:Tsource}) is modified to \cite{ref:MGCAMB} 
\begin{eqnarray}
\dot{\eta}+\ddot{\alpha} &=& \frac{\kappa}{2k^2} \left\{  - \left[(\gamma+1) (\dot{\rho} \Delta + \rho \dot{\Delta}) + \gamma \frac{3}{2}(\rho +P)\dot{\sigma} + \gamma \frac{3}{2} (\dot{\rho}+\dot{P})
 \sigma \right]\right. \nonumber\\
&+& \left.\dot{\gamma} \mu \left[ {(\rho \Delta)} + \frac{3}{2}(\rho +P)\sigma \right] \right\}, \label{eq:MGISW} 
\end{eqnarray}
which is the time derivative of the summation of the Newtonian potentials
\begin{equation}
\Psi+\Phi=\dot{\alpha}+\eta,
\end{equation} 
that is modified by the $\mu$ and $\gamma$ terms through the variation of $\alpha$. The CMB TT source term from the tenor mode perturbation in terms of the synchronous gauge variables is given by \cite{Zaldarriaga:1996xe}
\begin{equation}
S^{(T)}_T = -\dot{h}e^{-\kappa}+g\tilde{\Psi},
\end{equation}
where $\tilde{\Psi}$ denotes the temperate and polarisation perturbations generated by gravitational waves. Here $\dot{h}$ is related to $\alpha$ by 
\begin{equation}
\dot{h}=2k^2\alpha-6\dot{\eta}.
\end{equation} 

Now we move to study the effects on the CMB TT and BB power spectrum as shown in Figure \ref{fig:CMBTTBB}, by fixing the relevant cosmological parameters to their mean values obtained by {\it Planck} group \cite{ref:Planck2013CP} and $r=0.2$ by BICEP2 group \cite{ref:BICEP22}, but varying the parameters contained in $\gamma$, $\mu$, $\xi$ and $c^2_{T}$ freely. When $\lambda_{i}=1$, $s_{i}=0$ ($i=\mu,\gamma,t$) and $c^2_{T}=1$ are respected, the standard $\Lambda$CDM cosmology is recovered. It is called the corresponding standard value. 

\begin{widetext}
\begin{center}
\begin{figure}[tbh]
\includegraphics[width=8.9cm]{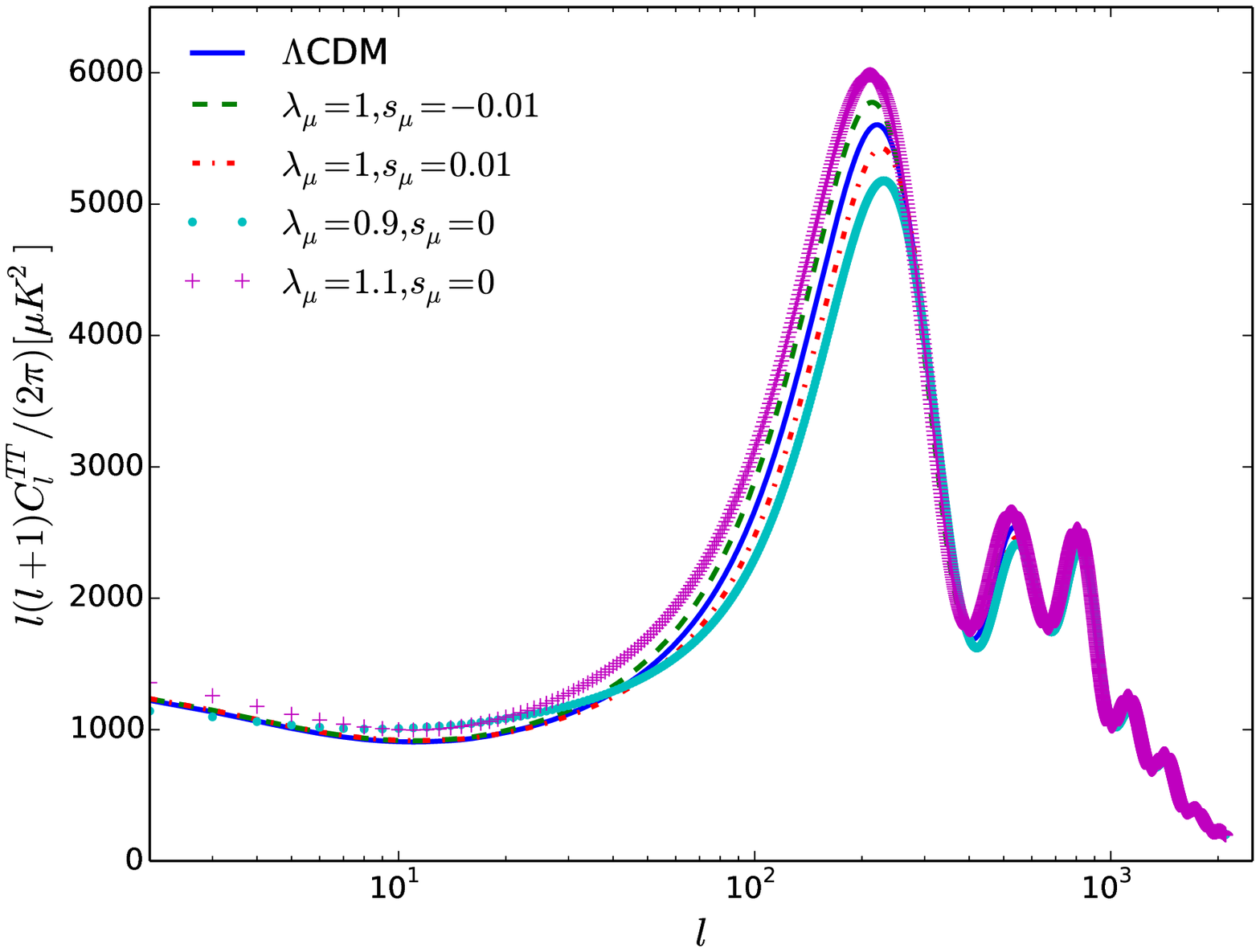}
\includegraphics[width=8.8cm]{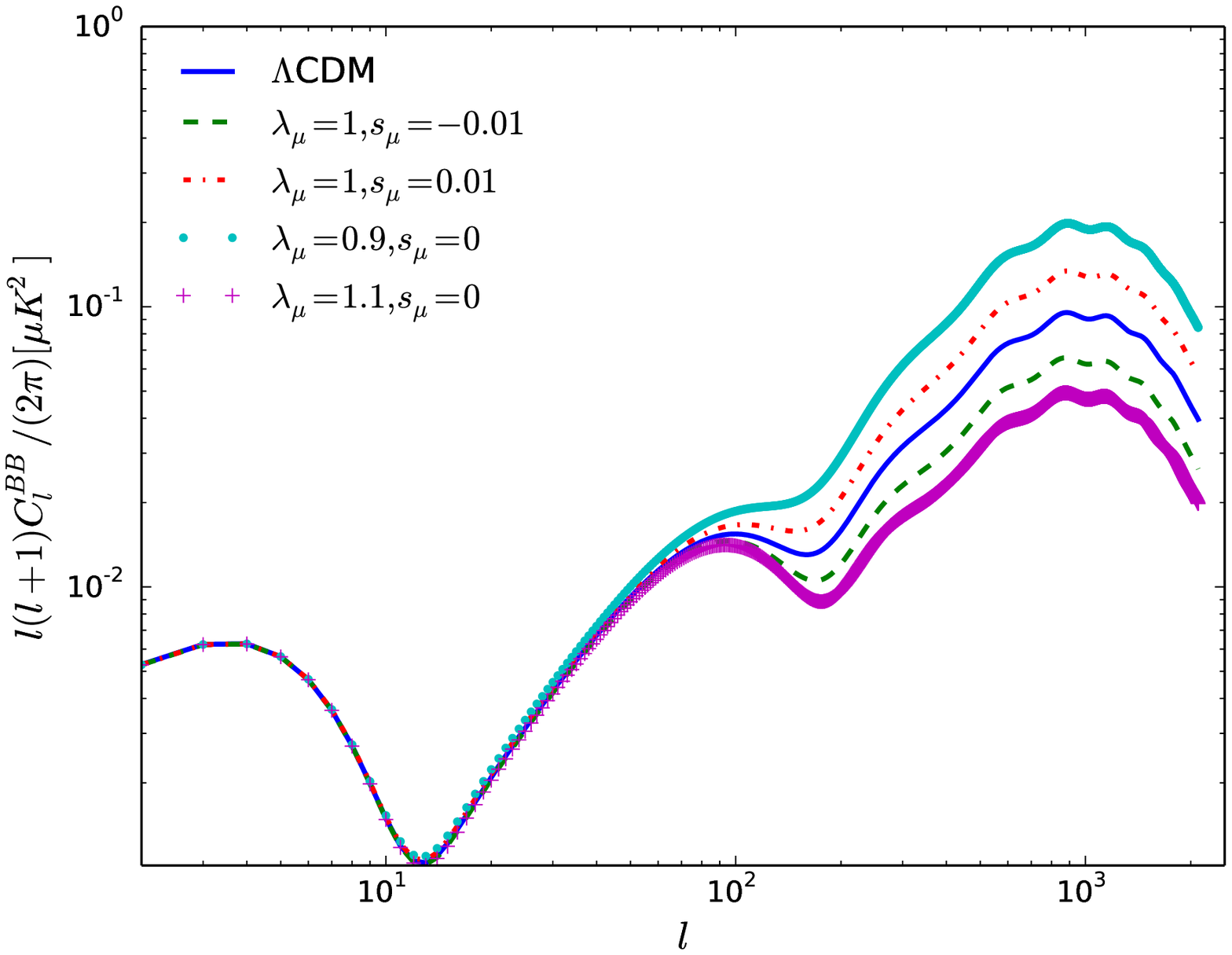}
\includegraphics[width=8.9cm]{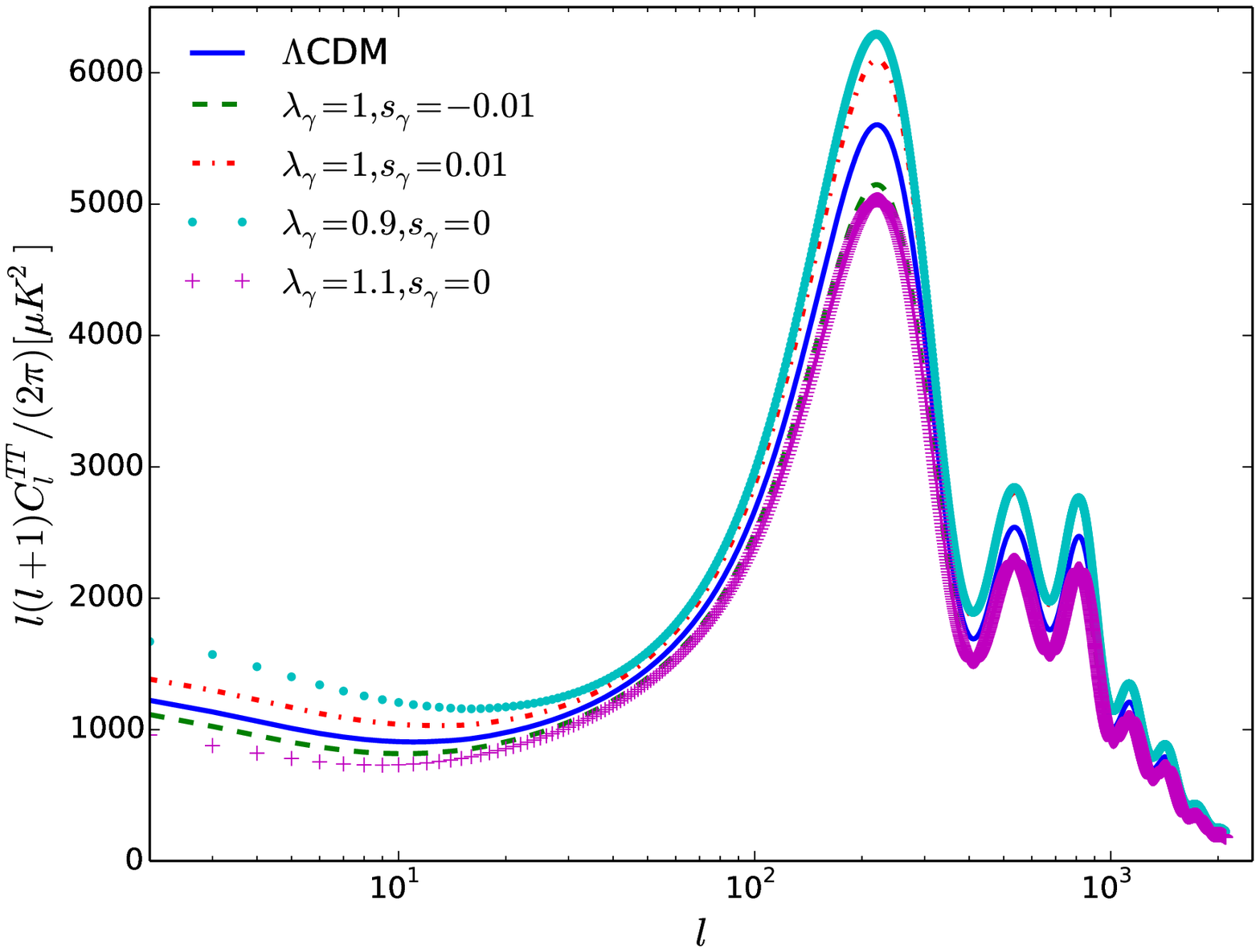}
\includegraphics[width=8.8cm]{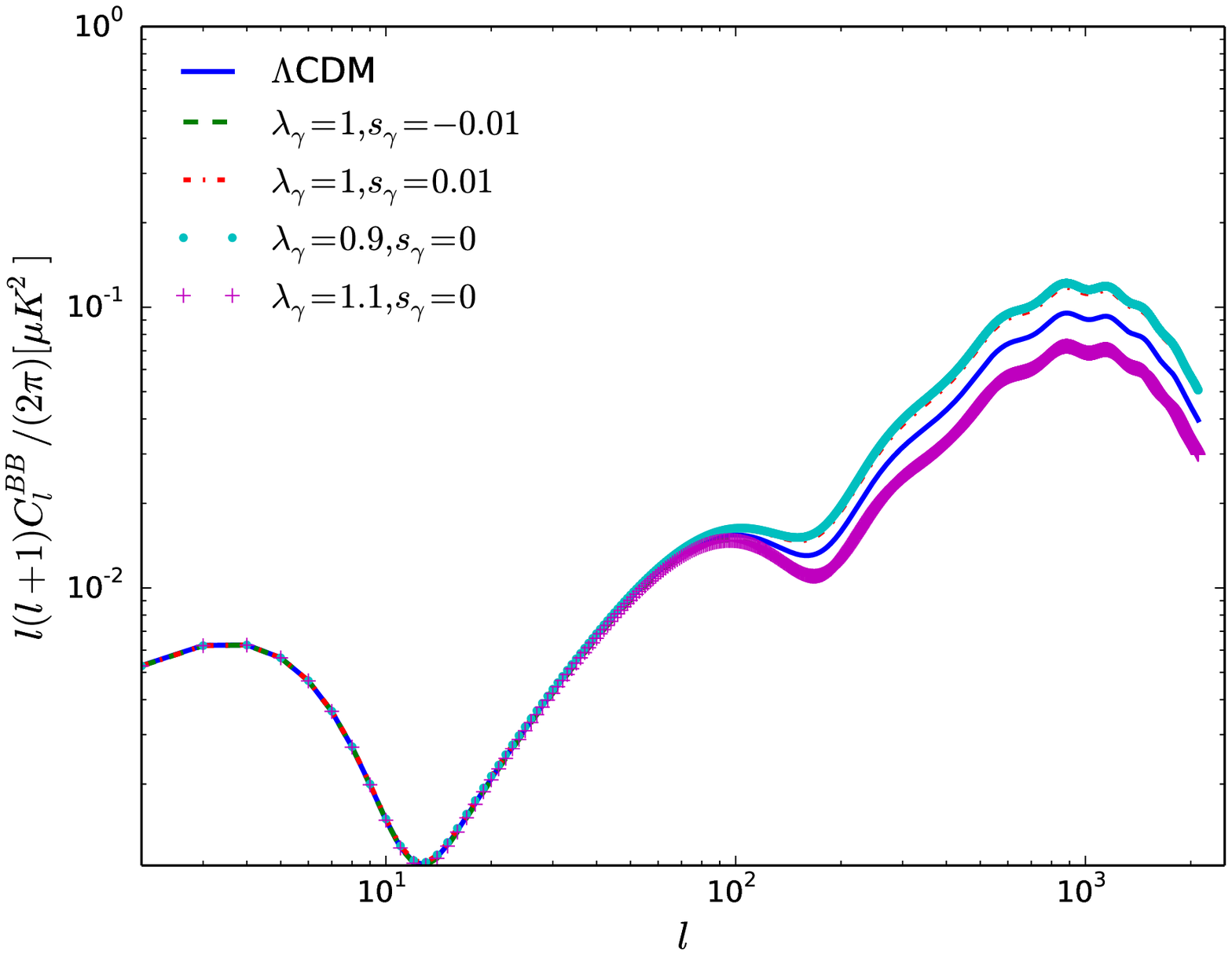}
\includegraphics[width=8.9cm]{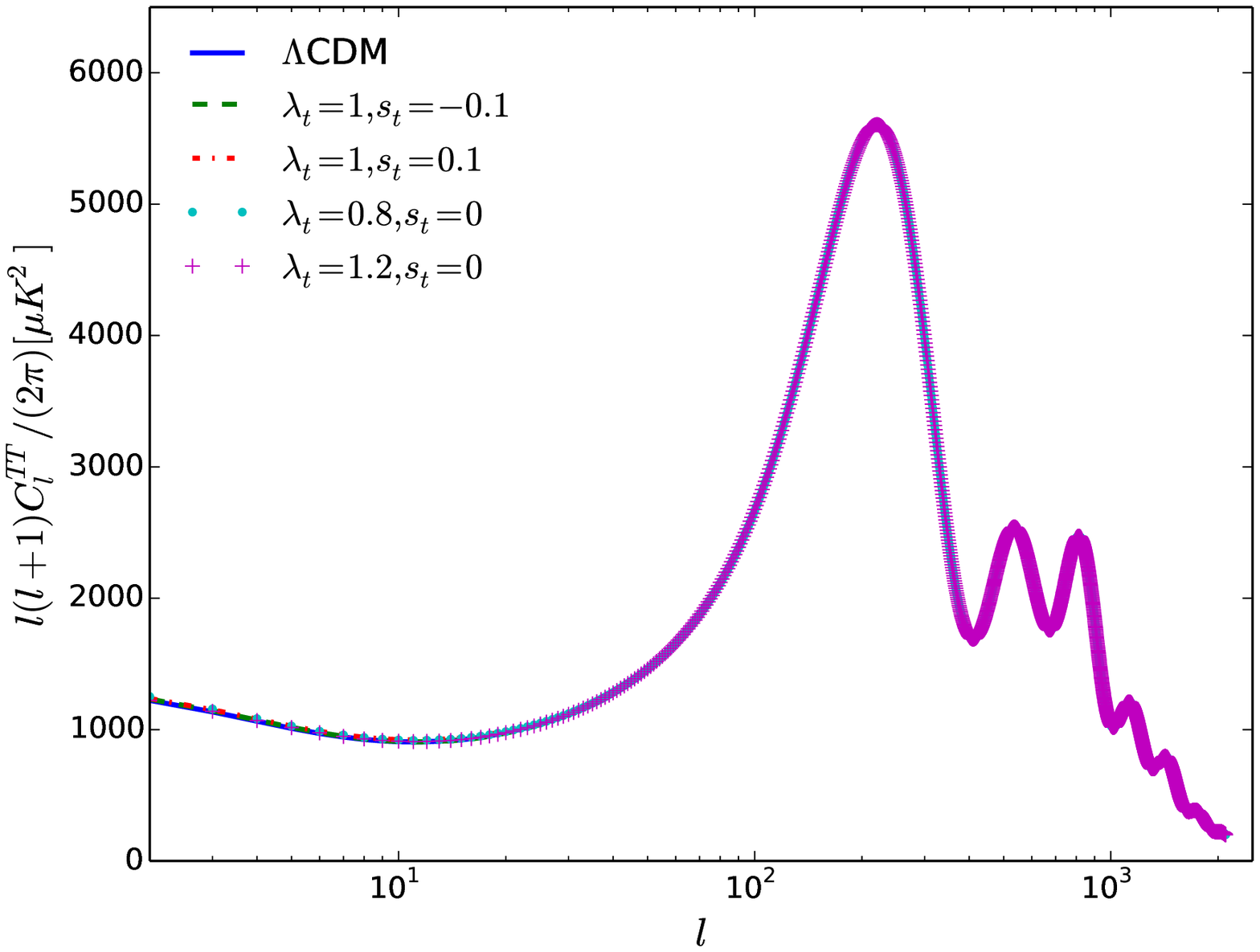}
\includegraphics[width=8.9cm]{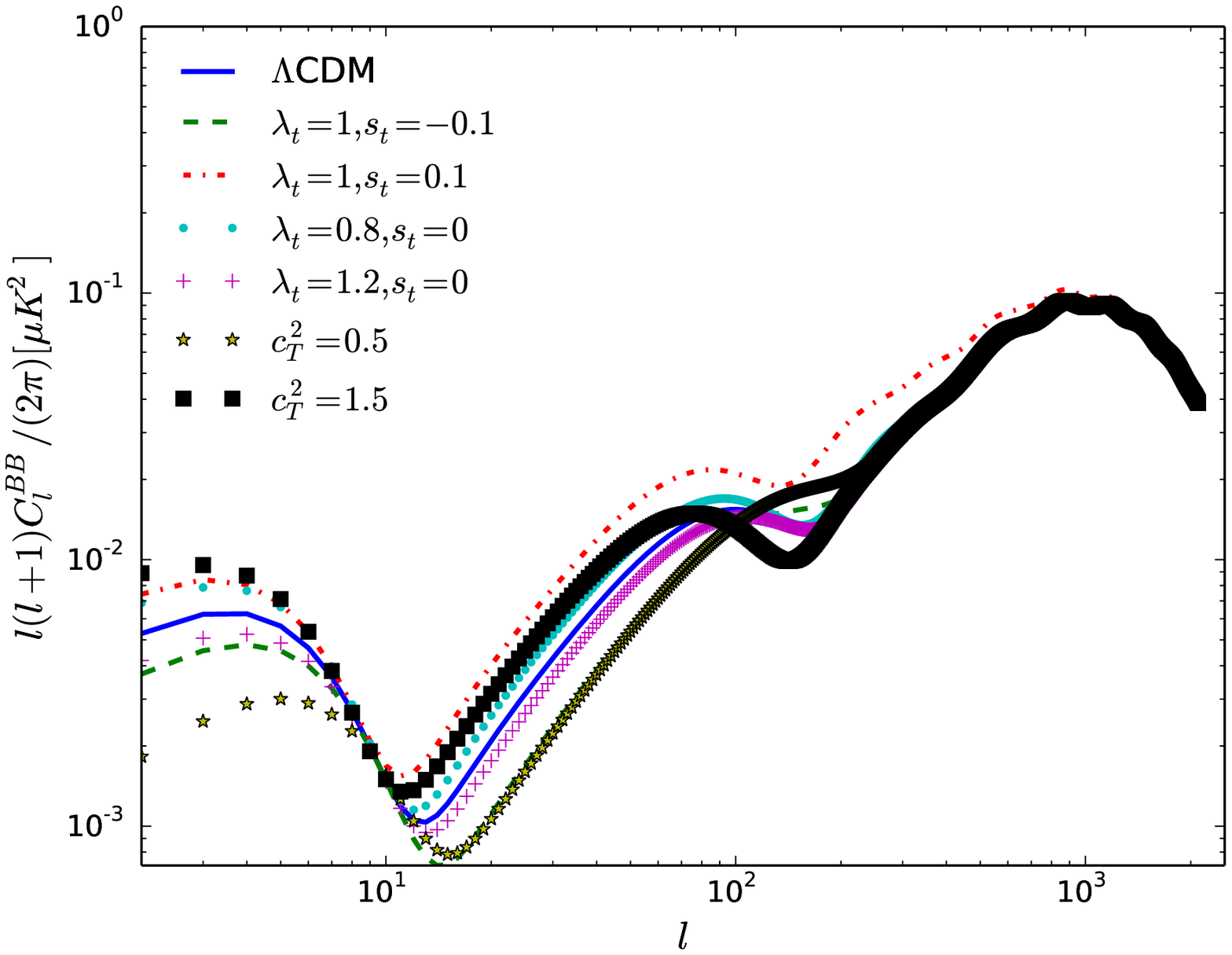}
\caption{The effects on the CMB TT (left panel) and BB (right panel) power spectrum arising from the variation of $\mu$ (top tow panels), $\gamma$ (middle two panels) and $\xi$ (bottom two panels) terms, where the relevant cosmological parameters are fixed to their mean values obtained by {\it Planck} group \cite{ref:Planck2013CP} and $r=0.2$ by BICEP2 group \cite{ref:BICEP22}. For every two panels, the other relevant MG parametrisation terms are fixed to their standard values.} \label{fig:CMBTTBB}
\end{figure}
\end{center}
\end{widetext}

In the top two panels of Figure \ref{fig:CMBTTBB}, we show the effects on the CMB TT (left panel) and BB (right panel) power spectrum resulting from the variation of the $\mu$ term, by fixing $\gamma\equiv 1$, $\xi\equiv 1$ and $c^2_{T}\equiv 1$. For the CMB TT power spectrum, since $\gamma\equiv 1$ is fixed, the ISW term is untouched (retained to the standard $\Lambda$CDM model). The change of the amplitude of CMB TT power spectrum is mainly caused by the integration of $\mu$ and $\dot{\mu}$ terms through the SW effect (the $g\dot{\alpha}$ and $\dot{g}\alpha$ terms) along the line of sight, i.e. at the range of $20<\ell<200$. Actually it is the result of the competition between $S^{(S)}_T$ and $S^{(T)}_T$ source terms. For the CMB BB power spectrum, the tensor perturbation $h$ is sourced by the $\mu$ term, therefore the amplitude of the CMB BB power spectrum is enlarged with the increase of $\mu$ through the lensing effects at multipole $l>100$ region. And this modification keeps the CMB BB power spectrum almost untouched at low multipole $\ell<10$, where the BB power spectrum is mainly dominated by the primordial gravitational waves.   

In the middle two panels of Figure \ref{fig:CMBTTBB}, the effects on CMB TT (left panel) and BB (right panel) power spectrum due to the variation of $\gamma$ term are shown. The term contributes not only to the SW effect but also to the early and late ISW effect. Therefore it makes an observable change at the low multipole $\ell <20$. This late ISW effect arising from the evolution of $\gamma$ which cannot be produced by the $\mu$ term. Similar to the $\mu$ term, the contribution to the CMB BB power spectrum mainly comes from the lensing effects, but $\gamma$ term makes its amplitude change along to the contrary direction. 

In the bottom two panels of Figure \ref{fig:CMBTTBB}, we show the effects on the CMB TT (left panel) and BB (right panel) power spectrum arising from the variation of the $\xi(k,a)$ term in the case of a fixed $c^2_{T}=1$. The $\xi(k,a)$ term only modifies the propagation of the primordial gravitational waves through the friction term $\dot{h}^{T}$, but keeps the CMB TT power spectrum untouched. Increasing the friction depresses the amplitude and make it move to the right direction. This effect happens at the low multipole $\ell<100$ and cannot be mimicked by the $\mu$ and $\xi$ terms which usually modify the relations between Newtonian potentials. Therefore, the CMB BB power spectrum at low multipole provides a unique character distinguishing MG from DE model in principle. In the right panel of Figure \ref{fig:CMBTTBB}, we also show the effects on the CMB BB power spectrum with a varying $c^2_{T}$. When $c^2_T$ varies for a MG model with fixed $\xi\equiv 1$, it can mimic the effect on the CMB BB power spectrum as that of $\xi$ term. That is the degeneracy between $\xi(a)$ and $c^2_{T}$ terms as shown in Fig. \ref{fig:CMBTTBB}. This confirms the results obtained in Ref. \cite{ref:MGmeature} and the reliability of our code. The degeneracy happens when $2\xi(k,a)\mathcal{H}\dot{h}^{T}_{ij}\sim c^2_Tk^2h^{T}_{ij}$ is respected. And this degeneracy makes it more difficult to detect MG from CMB BB power spectrum.   

However, in the above investigation, the primordial power spectrum is specified by a fixed tensor-to-scalar ratio $r=0.2$. But commonly the primordial power spectrum is characterised by the tensor-to-scalar ratio $r$, the tensor mode power spectrum index $n_{t}$ and its running $\alpha_{t}=d n_{t}/d\ln k$. So one should check whether the shift of the CMB BB power spectrum at the multipole $\ell < 100$ is caused by MG or modification of the propagation equation of gravitational waves. In doing so, we show the effects on the CMB TT and BB power spectrum arising from different values of $r$, $n_{t}$ and $\alpha_{t}$ in GR for the standard $\Lambda$CDM cosmology in Figure \ref{fig:LCDMTTBB}. Here we only focus on the CMB BB power spectrum. Large $r$ values increase the ratio of $A_{t}/A_{s}$ and make the whole BB power spectrum move along to the vertical direction at the multipole $\ell<100$, but retaining the shape. The index $n_{t}$ changes the amplitude and shape of the power spectrum simultaneously at the range $10<\ell<100$. The running $\alpha_t$ changes the shape at $\ell<10$ and has little effects to the power spectrum in the range $10<\ell<100$. Therefore careful choices of $r$, $n_{t}$ and $\alpha_{t}$ can mimic the effects arising from the MG $\xi$ term. Thus to distinguish MG from DE models, one still needs to understand the inflation very well. So in this work, we assume the inflation model is parametrised by $r$ only.

\begin{widetext}
\begin{center}
\begin{figure}[tbh]
\includegraphics[width=8.9cm]{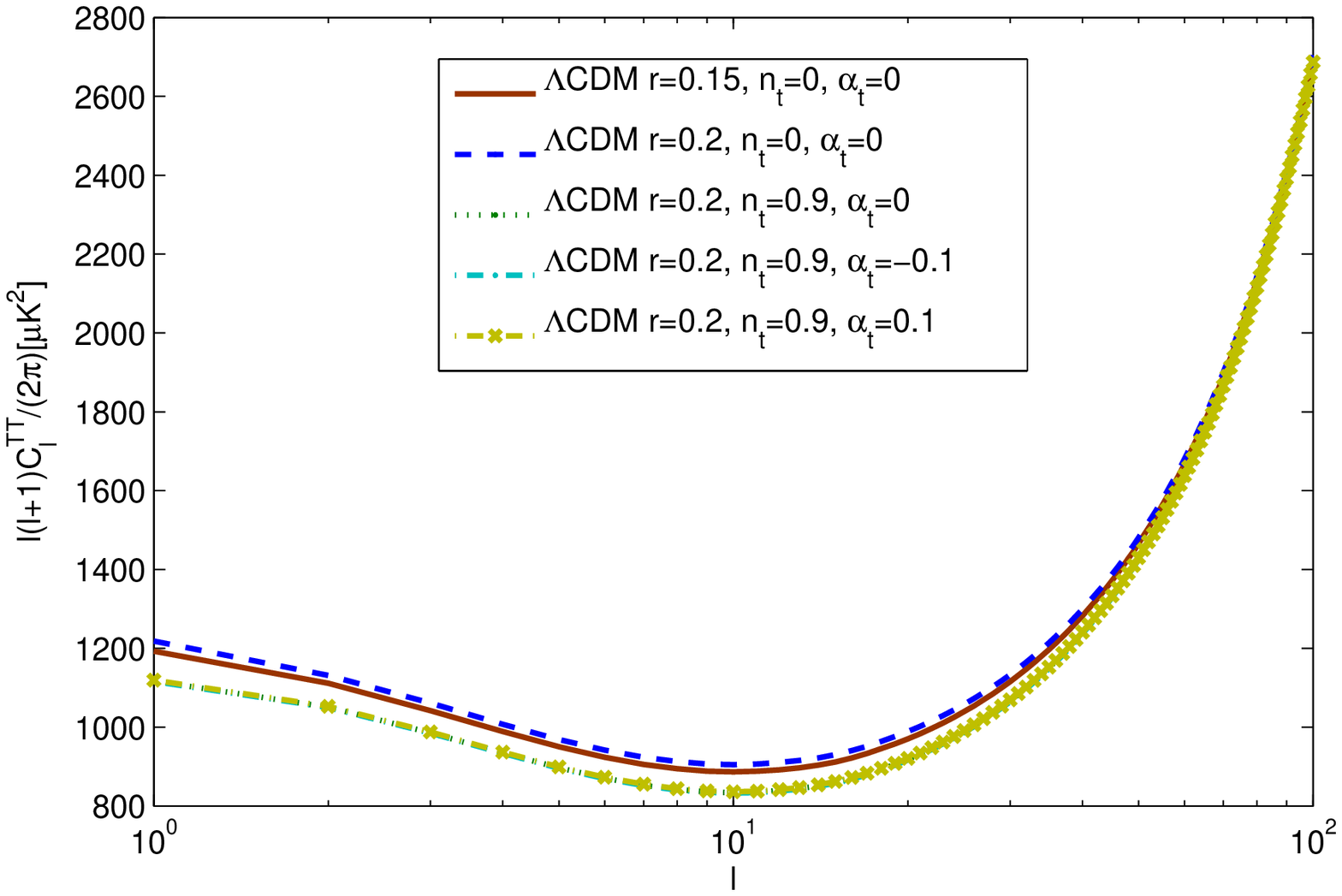}
\includegraphics[width=8.8cm]{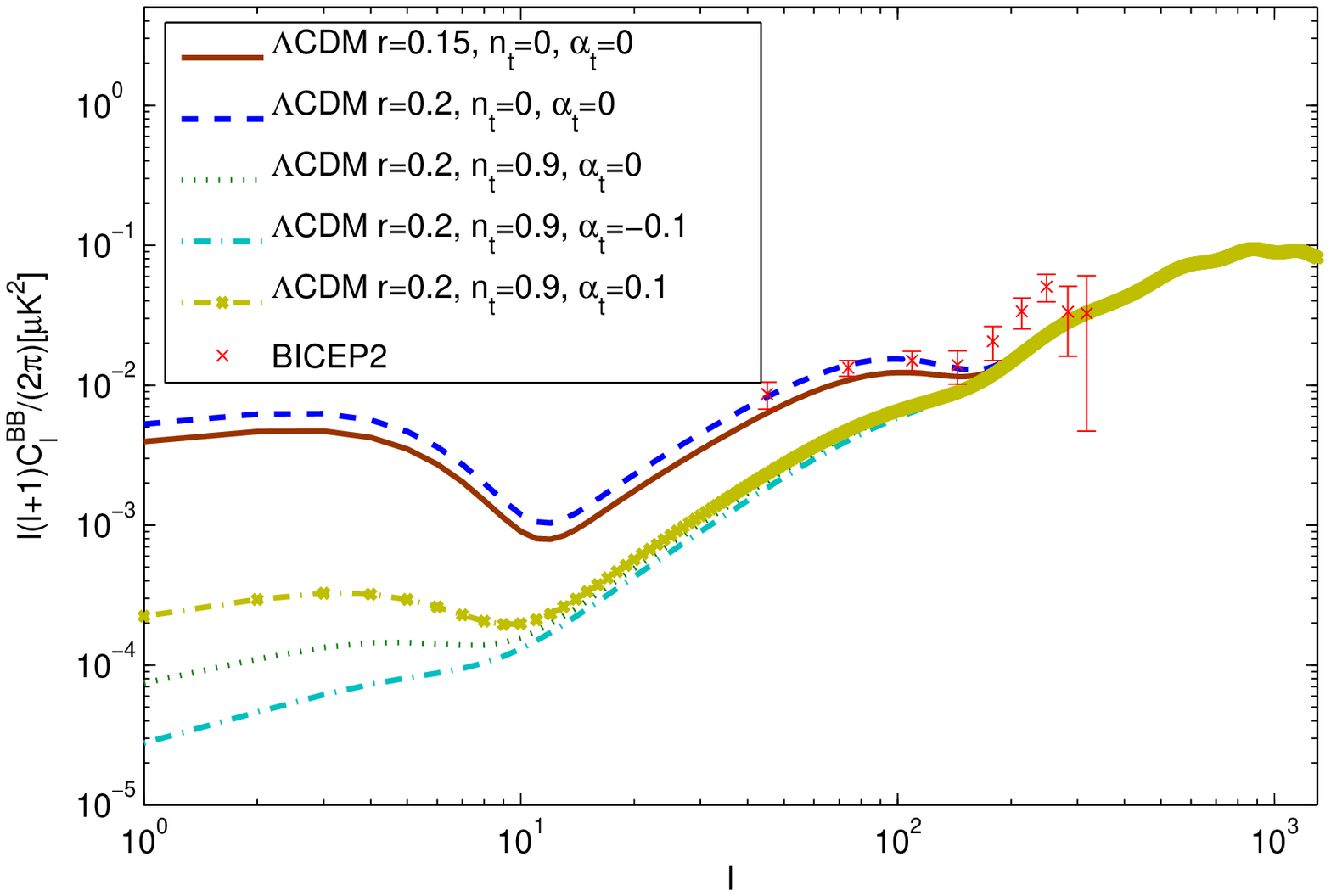}
\caption{The effects on the CMB TT and BB power spectrum with the variation of $r$, $n_{t}$ and $\alpha_{t}$, where the relevant cosmological parameters are fixed to their mean values obtained by {\it Planck} group \cite{ref:Planck2013CP}.}\label{fig:LCDMTTBB}
\end{figure}
\end{center}
\end{widetext}

\section{Data Set and Results} \label{sec:results}

In this section, we probe the signal of MG parametrised by $\mu$, $\gamma$ and $\xi$ terms by using the currently available cosmic observations which are summarised in the fowling,  based on the assumption that the inflation model is well understood and parametrised by $r$ only, here the $c^2_{T}$ is fixed to its standard value $1$:  

(i) The newly released BICEP2 CMB B-mode data \cite{ref:BICEP21,ref:BICEP22}. It will be denoted by BICEP2. Although the BICEP2 data has been confirmed as dust polarisation recently. We will still use these data points, then it can be taken as a test with significant primordial gravitational waves signals in the future.  

(ii) The full information of CMB which include the recently released {\it Planck} data sets which include the high-l TT likelihood ({\it CAMSpec}) up to a maximum multipole number of $l_{max}=2500$ from $l=50$, the low-l TT likelihood ({\it lowl}) up to $l=49$ and the low-l TE, EE, BB likelihood up to $l=32$ from WMAP9, the data sets are available on line \cite{ref:Planckdata}. This dat set combination will be denoted by P+W.

(iii) For the BAO data points as 'standard ruler', we use the measured ratio of $D_V/r_s$, where $r_s$ is the co-moving sound horizon scale at the recombination epoch, $D_V$ is the 'volume distance' which is defined as
\begin{equation}
D_V(z)=[(1+z)^2D^2_A(z)cz/H(z)]^{1/3},
\end{equation}
where $D_A$ is the angular diameter distance. The BAO data include $D_V(0.106) = 456\pm 27$ [Mpc] from 6dF Galaxy Redshift Survey \cite{ref:BAO6dF}; $D_V(0.35)/r_s = 8.88\pm 0.17$ from SDSS DR7 data \cite{ref:BAOsdssdr7}; $D_V(0.57)/r_s = 13.62\pm 0.22$ from BOSS DR9 data \cite{ref:sdssdr9}. This data set combination will be denoted by BAO.

(iv) The ten $f\sigma_8(z)$ data points from the redshift space distortion (RSD) are used, they are summarized as in Table \ref{tab:fsigma8data}.
\begin{center}
\begin{table}[tbh]
\begin{tabular}{cccl}
\hline\hline 
$\sharp$ & z & $f\sigma_8(z)$ & Survey and Refs \\ \hline
$1$ & $0.067$ & $0.42\pm0.06$ & 6dFGRS~(2012) \cite{ref:fsigma85-Reid2012}\\
$2$ & $0.17$ & $0.51\pm0.06$ & 2dFGRS~(2004) \cite{ref:fsigma81-Percival2004}\\
$3$ & $0.22$ & $0.42\pm0.07$ & WiggleZ~(2011) \cite{ref:fsigma82-Blake2011}\\
$4$ & $0.25$ & $0.39\pm0.05$ & SDSS~LRG~(2011) \cite{ref:fsigma83-Samushia2012}\\
$5$ & $0.37$ & $0.43\pm0.04$ & SDSS~LRG~(2011) \cite{ref:fsigma83-Samushia2012}\\
$6$ & $0.41$ & $0.45\pm0.04$ & WiggleZ~(2011) \cite{ref:fsigma82-Blake2011}\\
$7$ & $0.57$ & $0.43\pm0.03$ & BOSS~CMASS~(2012) \cite{ref:fsigma84-Reid2012}\\
$8$ & $0.60$ & $0.43\pm0.04$ & WiggleZ~(2011) \cite{ref:fsigma82-Blake2011}\\
$9$ & $0.78$ & $0.38\pm0.04$ & WiggleZ~(2011) \cite{ref:fsigma82-Blake2011}\\
$10$ & $0.80$ & $0.47\pm0.08$ & VIPERS~(2013) \cite{ref:fsigma86-Torre2013}\\
\hline\hline
\end{tabular}
\caption{The data points of $f\sigma_8(z)$ measured from RSD with the survey references.}
\label{tab:fsigma8data}
\end{table}
\end{center}
The scale dependence of the growth rate $f=d\ln \Delta_m/d\ln a$ in a gravity theory beyond GR at linear scale was reported in Refs. \cite{ref:scale,ref:frXu}, where $\Delta_m=\delta_m+3\mathcal{H}(1+w_m)\theta_m/k^2$. Thus the product $f(z,k)$ and $\sigma_8(z)$, i.e. $f\sigma_8(z,k)$, depends on the scale $k$ obviously, since $\sigma_8(z)$ is the function only  of redshift $z$. To remove this explicit scale dependence of $f\sigma_8(z)$, we should define it in theory as
\begin{equation}
f\sigma_8(z)=\frac{d\sigma_8(z)}{d\ln a},\label{ref:fsigma8def}
\end{equation}    
which is scale independent for any gravity theory and cosmological model. And the conventional definition is recovered for GR. Here we would like to warn the reader that the observed values of $f\sigma_8(z)$ are obtained based on the standard $\Lambda$CDM model, and are still unavailable for MG. With the observations on the Figure 11 in Ref. \cite{ref:frRSD}, say in the regime $k<0.1 h/\text{Mpc}$ at $z=0$ for $|f_{R0}|=10^{-4}$, the linear theory prediction for the growth rate almost matches the $N$-body simulation results for the $f(R)$ model, but deviates to the GR ones about $20\%$. Therefore, we naively assume that the underlying complication (including the scale dependence of the growth rate $f(z,k)$) can enlarge the error bars listed in Table \ref{tab:fsigma8data} to $20\%$, when the model parameter space is constrained. Therefore, we can take it as a preliminary results from RSD constraint.   

(v) The consistence of $\Omega_m$ between Ia supernovae and {\it Planck} 2013 was shown by SDSS-II/SNLS3 joint light-curve analysis, for the details please see \cite{ref:SNJLA}.   

(vi) The present Hubble parameter $H_0 = 73.8\pm 2.4$ [$\text{km s}^{-1} \text{Mpc}^{-1}$] from HST \cite{ref:HST} is used.

We perform a global fitting to the model parameter space
\begin{equation}
P=\{\Omega_b h^2,\Omega_c h^2,  100\theta_{MC}, \tau, {\rm{ln}}(10^{10} A_s),n_s, r, \lambda_{\mu},s_{\mu},\lambda_{\gamma},s_{\gamma},\lambda_{t}, s_{t}\},
\end{equation}
on the {\it Computing Cluster for Cosmos} by using the publicly available package {\bf CosmoMC} \cite{ref:MCMC}. The priors for the model parameters are shown in the second column of Table \ref{tab:results}. The running was stopped when the Gelman \& Rubin $R-1$ parameter $R-1 \sim 0.02$ is satisfied; that guarantees the accurate confidence limits. The obtained results are shown in Table \ref{tab:results} for the data combinations:  {\it Planck} 2013, WMAP9, BAO, BICEP2, JLA, HST and RSD. The obtained contour plots for the interested model parameters are shown in Figure \ref{fig:contour}. 

\begingroup                                                                                                                     
\squeezetable                                                                                                                   
\begin{center}                                                                                                                  
\begin{table}                                                                                                                   
\begin{tabular}{cccc}                                                                                                            
\hline\hline                                                                                                                    
Parameters & Priors & Mean with errors & Best fit \\ \hline
$\Omega_b h^2$ & $[0.005,0.1]$ & $0.02201_{-0.00038}^{+0.00038}$ & $    0.02200$\\
$\Omega_c h^2$ & $[0.001,0.99]$ & $0.1177_{-0.0017}^{+0.0017}$ & $0.1161$\\
$100\theta_{MC}$ & $[0.5,10]$  & $1.0436_{-0.0013}^{+0.0013}$ & $1.0431$\\
$\tau$ & $[0.01,0.81]$ & $0.089_{-0.014}^{+0.013}$ & $0.091$\\
${\rm{ln}}(10^{10} A_s)$ & $[2.7,4]$ & $3.113_{-0.088}^{+0.089}$ & $3.135$\\
$n_s$ & $[0.9,1.1]$ & $0.958_{-0.042}^{+0.026}$ & $0.970$\\
$r$ & $[0,1]$ & $0.045_{-0.038}^{+0.008}$ & $0.058$\\
$\lambda_{\mu}$ & $[0,2]$ & $1.030_{-0.023}^{+0.023}$ & $1.021$\\
$s_{\mu}$ & $[-1,1]$ & $0.0054_{-0.0037}^{+0.0037}$ & $0.0042$\\
$\lambda_{\gamma}$ & $[0,2]$ & $1.11_{-0.15}^{+0.16}$ & $1.07$\\
$s_{\gamma}$ & $[-1,1]$ & $0.007_{-0.019}^{+0.027}$ & $0.002$\\
$\lambda_{t}$ & $[0,2]$ & $-$ & $0.95$\\
$s_{t}$ & $[-1,1]$ & $0.62_{-0.73}^{+0.31}$ & $0.19$\\
\hline
$H_0$ & $73.8 \pm 2.4$ & $68.78_{-0.76}^{+0.77}$ & $69.15$\\
$\Omega_\Lambda$ & $...$ & $0.7032_{-0.0094}^{+0.0095}$ & $0.7098$\\
$\Omega_m$ & $...$ & $0.2968_{-0.0095}^{+0.0094}$ & $0.2902$\\
$\sigma_8$ & $...$ & $0.798_{-0.021}^{+0.021}$ & $0.803$\\
$z_{\rm re}$ & $...$ & $10.95_{-1.13}^{+1.14}$ & $11.12$\\
${\rm{Age}}/{\rm{Gyr}}$ & $...$ & $13.745_{-0.046}^{+0.046}$ & $13.753$\\
\hline\hline                                                                                                                    
\end{tabular}                                                                                                                   
\caption{The mean values with $1\sigma$ errors and the best fit values of the model parameters and the derived cosmological parameters, where the {\it Planck} 2013, WMAP9, BAO, BICEP2, JLA, HST and RSD data sets were used. '$-$' denotes the one which is not well constrained.}\label{tab:results}                                                                                                   
\end{table}                                                                                                                     
\end{center}                                                                                                                    
\endgroup 

\begin{widetext}
\begin{center}
\begin{figure}[tbh]
\includegraphics[width=16.5cm]{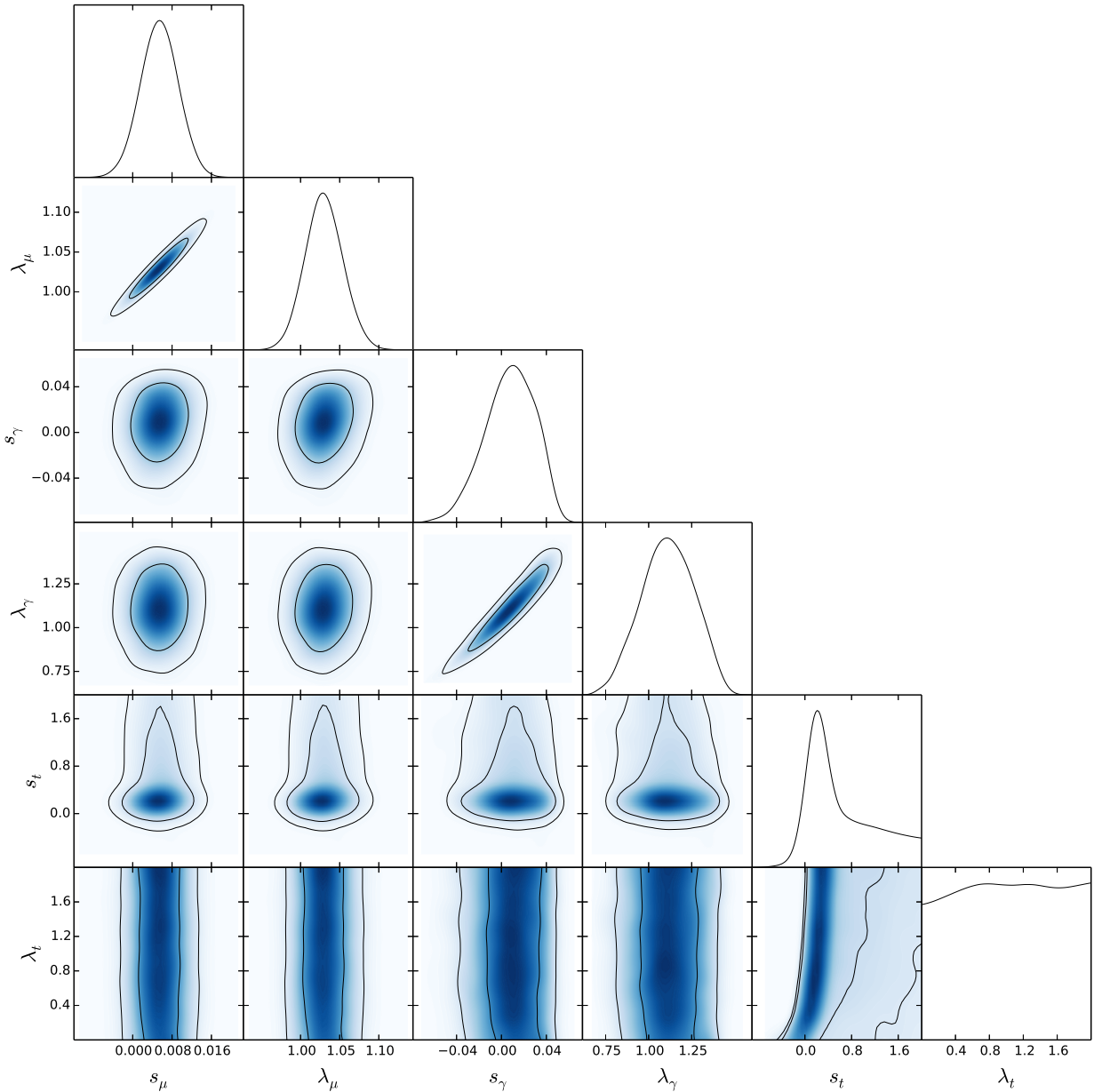}
\caption{The 1D marginalised distribution and 2D contours for interested model parameters with $68\%$ C.L., $95\%$ C.L. by using the {\it Planck} 2013, WMAP9, BAO, BICEP2, JLA, HST and RSD data sets.}\label{fig:contour}
\end{figure}
\end{center}
\end{widetext}

One can clearly see that no significant deviation from GR was detected for the scaler perturbations in $2\sigma$ regions. Although, for the scalar perturbations modelling the modification to the Poisson equation, slight deviations to the standard values in $1\sigma$ regions for model parameters $\lambda_{\mu}=1.030_{-0.023}^{+0.023}$ and $s_{\mu}=0.0054_{-0.0037}^{+0.0037}$ are shown. This tension was also reported in the {\it Planck} 2015 paper for dark energy and modified gravity \cite{ref:planck2015demg}. And this tension can be reconciled by including the CMB lensing \cite{ref:planck2015demg}. It is interesting to show the correlations for model parameters pairs $\lambda_\mu-s_\mu$, and $\lambda_\gamma- s_\gamma$. The uncorrelation between the $\mu$ and $\gamma$ terms implies that they have different sources and cannot mimic each other.   

\begin{center}
\begin{figure}[tbh]
\includegraphics[width=8.5cm]{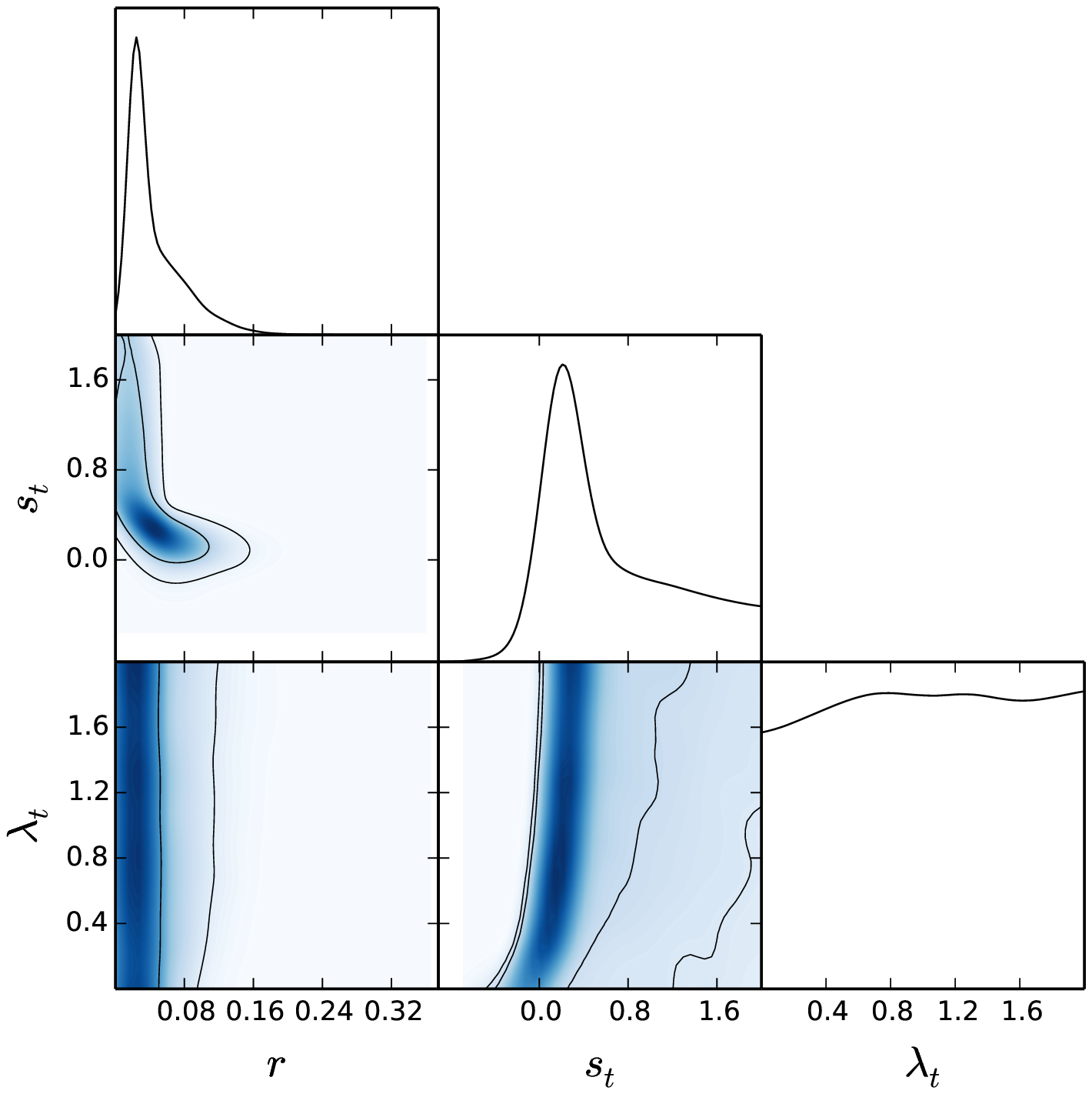}
\caption{The same as Figure \ref{fig:contour}, but for the model parameters $r$, $\lambda_t$ and $s_t$.}\label{fig:r}
\end{figure}
\end{center}

For the tensor perturbations, as shown in Figure \ref{fig:r}, one can see the anti-correlation between model parameters $s_t$ and $r$. This can explain the small values of $r$. The model parameter $\lambda_t$ cannot be well constrained due to the lack of data points below $l<10$. And a detection of the deviation to GR in this region is a tough task due to the domination by the cosmic variance.

In the whole global fitting process, we have fixed $c^2_{T}$ to its standard value $1$. If it is taken as another free model parameter, one cannot obtain a tight constraint to $c^2_{T}$ based on currently available data points due to the degeneracy to the $\xi$ term which is not well constraint as shown in Fig. \ref{fig:contour} and Table \ref{tab:results}. Therefore, the introduction of free $c^2_{T}$ will not change the main results of our analysis. 

\section{Conclusion} \label{sec:conclusion} 

In this paper,  we proposed a parameterised time dependent modification to the propagation of gravitational waves, since the scale dependence does not lead to a significantly small $\chi^2$ with respect to the scale-independent case \cite{ref:Planck2015DEMG}. Taking this specific form as a working example, we showed the effects to the CMB TT and BB power spectrum due to this kind of modification to GR by adopting different values of the model parameters. We also showed the possible degeneracy to the tensor mode power spectrum index $n_t$ and its running $\alpha_t$. Our analysis reveals that the modification to GR at tensor mode perturbations has effects on the CMB BB power spectrum at low multipole $l < 10$, i.e. the large scale, and keeps the shape of the CMB BB power spectrum. And the tensor mode power spectrum index $n_t$ and its running $\alpha_t$ have effects to CMB BB power spectrum in the range $l\in (1,100)$ and change the shape of the CMB BB power spectrum. It implies a precise data points below $l\sim 10$ can break this degeneracy between modification to GR and the power spectrum index and its running. However it is a tough task due to the domination by the cosmic variance in this region.  

We also used the currently available cosmic observational data sets, which include {\it Planck} 2013, WMAP9, BAO, BICEP2, JLA, HST and RSD, to detect the possible deviation to GR. The results were gathered in Table \ref{tab:results} and Figure \ref{fig:contour} and Figure \ref{fig:r}. We didn't find any significant deviation to GR in $2\sigma$ regions. But for the scalar perturbation part, we found the same tension as that reported in the {\it Planck} 2015 paper for dark energy and modified gravity \cite{ref:planck2015demg}: the slight deviations to the standard values in $1\sigma$ regions for model parameters $\lambda_{\mu}=1.030_{-0.023}^{+0.023}$ and $s_{\mu}=0.0054_{-0.0037}^{+0.0037}$. The uncorrelation between the $\mu$ and $\gamma$ terms implies that they have different sources and cannot mimic each other. For the tensor perturbation part, the model parameter $\lambda_t$ is not well constrained due to the lack of data points. The anticorrelation between model parameter $\lambda_t$ and $r$ was also shown. This anticorrelation explains the small values of $r$. 

Although in this paper, we have used the BICEP2 data points which are already confirmed as dust polarisation, the analysis on the effects to the CMB TT and BB power spectrum is still robust. And the correlation and anticorrelation of the model parameters are irrelevant to the BICEP2 data points.

\acknowledgements{The author thanks an anonymous referee for helpful improvement of this paper and thanks Dr. Bin Hu for useful discussion and ICTP for hospitality during the author's visit in ICTP. This work is supported in part by National Natural Science Foundation of China under Grant No. 11275035 (People's Republic of China), the Fundamental Research Funds for the Central Universities, and the Open Project Program of State Key Laboratory of Theoretical Physics, Institute of Theoretical Physics, Chinese Academy of Sciences No. Y4KF101CJ1 (People's Republic of China).}

\end{document}